\documentstyle[12pt]{article}
\begin{document}
\begin{center}

{\large \bf Effect of Earth's rotation on the trajectories of
free-fall bodies in Equivalence Principle Experiment}
\\

\vspace{8mm}
  C. G. Shao$^a$, Y. Z. Zhang$^{b,c}$, J. Luo$^a$ and Z. Z.
  Liu$^a$ \\
\vspace{4mm}
  {\footnotesize{\it
  $^a$Department of Physics, Huazhong University of Science and
        Technology,\\ Wuhan 430074, China \\
  $^b$CCAST(World Lab.), P. O. Box 8730, Beijing 100080, China \\
  $^c$Institute of Theoretical Physics, Chinese Academy of Sciences,
          P.O. Box 2735,\\ Beijing 100080, China }}

 \end{center}

\vspace{8mm}

\begin{abstract}
Owing to Earth's rotation a free-fall body would move in an
elliptical orbit rather than along a straight line forward to the
center of the Earth. In this paper on the basis of the theory for
spin-spin coupling between macroscopic rotating bodies we study
violation of the equivalence principle from long-distance
free-fall experiments by means of a rotating ball and a
non-rotating sell. For the free-fall time of 40 seconds, the
difference between the orbits of the two free-fall bodies is of
the order of $10^{-9}$cm which could be detected by a SQUID
magnetometer owing to such a magnetometer can be used to measure
displacements as small as $10^{-13}$ centimeters.\\

\noindent
  PACS number(s): 04.80.Cc
\end{abstract}

\vspace{10mm}
 As one of the fundamental hypotheses of Einstein's
general relativity, the equivalence principle (EP) has been tested
by many experiments [1--7]. Some different tests of EP for
gravitational self-energy [8] and spin-polarized macroscopic
objects [9,10] have been also reported. However, in all of the
experiments as well as the Satellite Test of the Equivalence
Principle (STEP), the Galileo Galilei (GG) space projects and the
MICROSCOPE space mission [11--13], it is non-rotating bodies that
are used. Recently, a theory of EP violation coming from the
spin-spin coupling between rotating macroscopic bodies was
developed [14,15] and preliminary experiments have been performed
[16--18].

In order to test with higher precision the possible violation of
the EP coming from the spin-spin coupling of
 rotating macroscopic bodies, some proposed long-distance free-fall
experiments by using a drop tower of one hundred meter high or a
drop capsule of 40 kilometer altitude have been suggested [15]. In
this type of experiments the drop test masses may be a
superconducting gyroscope consisting of an inner spinning
superconducting ball and an outer non-rotating shell, and the gap
between the inner ball and outer shell could be measured by means
of a superconducting quantum interference device (SQUID). The
superconducting gyro is suspended on the top of a vacuum capsule.
Cutting off the suspension wire and, at the same time, turning off
the power supply of the gyro when the vacuum capsule starts in
free-fall, both the inner spinning ball and the outer non-rotating
shell would then move freely in the vacuum capsule. If the Earth
did not rotate, the gyro would simply fall forward to the center
of the Earth. In this case, any relative shift between the inner
ball and the outer shell would arise only in the free-fall
direction rather than the horizontal one if the EP violates.
However, owing to the nonzero initial velocity of the gyro
associated with Earth's rotation the real trajectory of the gyro
would be an elliptical orbit rather than a straight line
perpendicular to Earth's surface. The expected relative
displacements between the inner ball and the outer shell would
then occur in both the vertical and horizontal directions. The aim
of this paper is to study the effect of Earth's rotation on the
free-fall of the gyroscope and give as functions of time the
expected vertical and horizontal relative shifts between the
rotating and non-rotating bodies. Consider the free-fall of the
above gyro, which at the initial time $t=0$ is at rest at height
$h$ over the Earth surface. According to the theory developed in
[15], the equation of motion for the rotating inner ball in
Earth's gravitational field takes the form
$$m\mathop {\bf r}\limits^{..}=-\frac{GmM}{r^3}{\bf r}
     -\kappa\frac{{\bf s} \cdot {\bf S}}{r^4}{\bf r}  \eqno(1)$$
to the first order of $\kappa$, where $m$ and $M$ are the masses
of the test body and the Earth, respectively, ${\bf s}$ and ${\bf
S}$ are the corresponding spin angular momentums, $\kappa$ is an
arbitrary parameter related to the spin-spin coupling strength and
 the possible violations of the mass-velocity relation and
equivalence principle. Solution of Eq. (1) to the first order of
$\delta$ is [15]
$$r=\frac{(1-\delta)p}{1+e {\rm cos}(\sqrt{1-\delta}\phi +\pi)},
 \eqno(2a)$$
and
  $$\delta =\kappa\frac{{\bf s}\cdot {\bf S}}
{GMm}\frac{1}{p},                      \eqno(2b)$$ where $p$ and
$e$ are integration constants. A nonzero value of $\delta$ would
give perihelion precession of planet's orbits in solar system and
also violate the equivalence principle. The observations of the
perihelion precession establish the upper limit on the parameter
$\kappa\leq 3.4 \times 10^{-31}{\rm gram}^{-1}$(see [15]). A
non-rotating and comoving coordinate system of Earth's center is
regarded as the inertial frame ($r, \theta$). Owing to Earth's
rotation, the free-fall gyro at the initial position $\left. {\bf
r} \right|_{t = 0}  = (R + h){\bf e}_r$ would have the initial
horizontal velocity $\left. {\bf v} \right|_{t = 0} = {\bf
e}_\theta \omega (R + h)\sin \alpha$ where $R$ is Earth's radius,
$\omega$ is the angular velocity of Earth's rotation, $\alpha$ is
the colatitude of the gyro, ${\bf e}_r$ and ${\bf e_\theta}$ are
the unit basic vectors in the inertial frame ($r, \theta$). The
integration constants then become
   $$p=\frac{{\omega ^2(R + h)^4\sin^2\alpha}}{{GM}},  \eqno(2c)$$
    $$ e = 1 - \frac{p}{{R + h}}(1 - \delta ).         \eqno(2d)$$
Denote the eccentric anomaly $E$ which can be used for describing
the angle $\theta$:
$$\tan\frac{{\sqrt{1-\delta}\theta}}{2}=\sqrt{\frac{{1-e}}
{{1+e}}}\tan\frac{E}{2}.      \eqno (2e)$$
  $E$ is determined by the Kepler equation
$$E+e\sin E=nt, ~~~~ n=\sqrt{GM/b^3}, ~~~~ b=(R+h)/(1+e).\eqno(2f)$$
Since higher order in $\delta$ has been ignored, the following
expansions can be made:
$$ r(t) = r'(t) + \Delta r(t), ~~~~~~ \theta(t) = \theta'(t)
 + \Delta \theta(t), $$
 $$E(t) = E'(t) + \Delta E(t), ~~~~~~ e = e'+ \Delta e, ~~~~~~ n = n' +
 \Delta n,                                           \eqno(3)$$
where the quantities with the prime are of zero-order in $\delta$,
and the $\Delta$ terms are proportional to $\delta$ (i.e. the
first order in $\delta$). Putting the expansions into Eqs. (2a, d,
e, f), we have the zero-order equations,
$$r' = \frac{p}{{1 +e'\cos (\theta ' + \pi )}},    \eqno(4a) $$
$$\tan\frac{{\theta '}}{2}=\sqrt{\frac{{1-e'}}{{1+e'}}}
\tan\frac{{E'}}{2},                               \eqno(4b)$$

$$E' +e'\sin E' =n't,                              \eqno(4c)$$
$$e'=1-p/(R+h), ~~~~n' =\sqrt{GM/b^{'3} },
~~~ b' =(R+h)/(1+e'),                                \eqno(4d)$$
and the first-order ones,
$$ \Delta r(t) = \frac{{ - p\delta (1 + e'\cos E')}}{{1 - e'^2 }} +
 \frac{{p\cos E'\Delta e - pe'\Delta E\sin E'}}{{1 - e'^2 }}
  + \frac{{2pe'(1 + e'\cos E')\Delta e}}{{(1 - e'^2 )^2 }},
                                                      \eqno(5a)$$
$$\Delta \theta (t) = \frac{{\Delta E - \Delta E e' \cos \theta '}}
{{\sqrt {1 - e'^2 } }} - \frac{{\sin E'\Delta e}}{{\sqrt {1 - e'^2 }
(1 + e\cos E')}} + \frac{{\delta \theta '}}{2},
                                                     \eqno(5b)$$
$$\Delta E(t) = \frac{{\Delta nt - \Delta e\sin E'}}{{1 + e'\cos E'}},
                                                       \eqno(5c)$$
$$\Delta e = \frac{{p\delta }}{{R + h}}, ~~~~\Delta n =
\frac{{3n\Delta e}}{{2(1 + e')}},
                                                   \eqno(5d)$$
where $p$ is given by Eq. (2c). For convenance of numerical
calculation we rewrite Eqs. (5a)--(5c) as
 $$\frac{{\Delta r(t)}}{\delta } = \frac{p}{{1 + e'}}\left(
  {\frac{{\sin ^2 E' -(1 - e')(1-\cos E')}}{{(1 + e')(1 + e'\cos
  E')}} - \frac{{3n'te'\sin E'}}{{2(1 + e')(1 + e'\cos E')}}}
  \right),                                          \eqno(6a)$$
 $$\frac{{\Delta\theta(t)}}{\delta}=\frac{{1-e'\cos \theta '}}
 {{\sqrt {1 - e'^2 } (1 + e\cos E')}}\frac{{3n't(1 - e')}}{{2(1
 + e')}} - \frac{{(2 - e'\cos \theta ')\sin E'(1 - e')}}{{\sqrt {1
 - e'^2 } (1 + e\cos E')}} + \frac{{\theta '}}{2},
                                                 \eqno(6b)$$
where Eqs. (5c) and (5d) have been used.

We now denote the Cartesian coordinate system $(X,Z)$ in the
orbital plane $(r, \theta)$ by
  $$X=r\sin\theta, ~~~~~~ Z=r\cos\theta,       \eqno(7)$$
where $Z$-axis is in the vertical direction of Earth's surface
 and $X$-axis is in the horizontal plane.
 The differences in unit of $\delta$ between the coordinates of the
two orbits in the frame $(X, Z)$ are
 $$\frac{\Delta Z}{\delta} = (r\cos\theta -r'\cos\theta')/\delta =
    \frac{\Delta r}{\delta}\cos\theta' -\frac{\Delta \theta}{\delta}
    r'\sin\theta',                                \eqno(8a)$$
 $$\frac{\Delta X}{\delta} = (r\sin\theta -r'\sin\theta')/\delta =
    \frac{\Delta r}{\delta}\sin\theta' +\frac{\Delta \theta}{\delta}
    r'\cos\theta'.                                \eqno(8b)$$

The zero-order equations (4a)-(4d) just describe the free-fall of
the non-rotating outer shell which has the same initial conditions
with the rotating inner ball, i.e., at the initial time $t=0$,
$r'(0)=r(0)=R+h, \theta (0)=\theta'(0)=0$ and $v'(0)=v(0)=\omega
(R+h)\sin\alpha$. Thus at the initial time the first order
quantities all vanish: $\Delta r(0)= \Delta\theta(0)=0$.
 Solving the equations (4), (6) and (8) we could get the
 differences in unit of $\delta$ for the orbits of the
 inner ball and outer shell, which are shown in Fig. 1.
\input{epsf.sty}
\begin{figure}
\begin{center}
  \leavevmode
 \epsfysize=2.5in
  \epsfbox{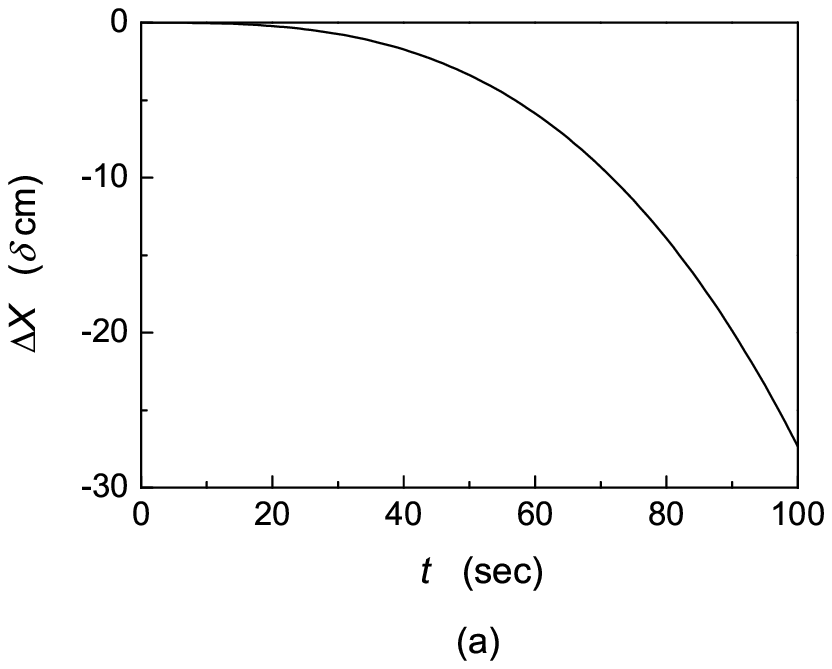}
 \epsfysize=2.5in
  \epsfbox{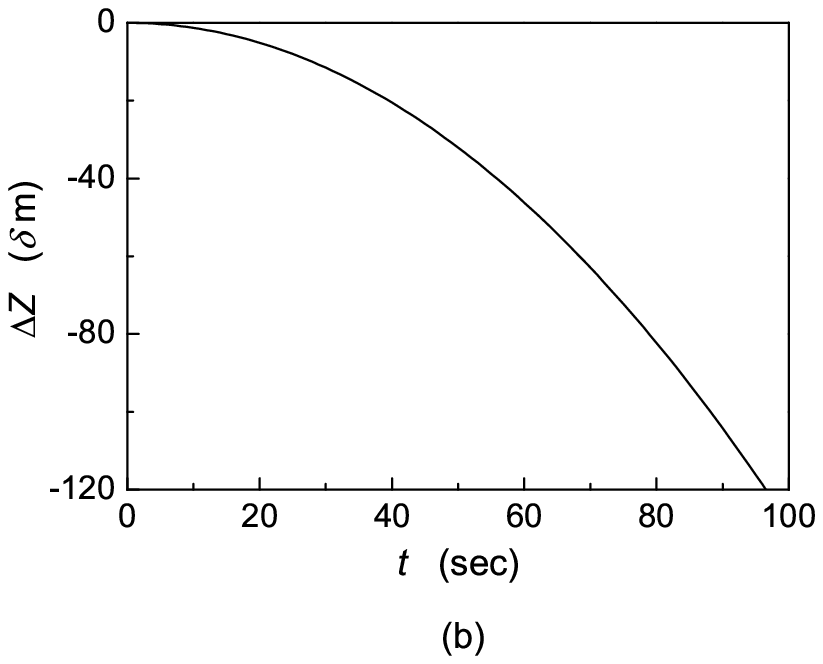}
\end{center}

 {\footnotesize Fig. 1. (a) and (b) show, respectively, $\Delta X$ and $\Delta Z$
  as functions of time $t$, where  $h=4\times 10^4$ m, $R=6.38\times
  10^6$ m and $a=\pi /3$ are used. The time $t$ is in unit of second.
   $\Delta X$ and $\Delta Z$ are, respectively, in units of
   $\delta\times$1cm and $\delta\times$1m.}
%\label{Fig1}
\end{figure}

We now consider acceleration of the rotating inner ball. From Eq.
(1) we have
  $$a_{Z}=-\frac{GM}{r^2}\cos\theta
     \left(1+\frac{p\delta}{r}\right),              \eqno(9a)$$
  $$a_{X}=-\frac{GmM}{r^2}\sin\theta
     \left(1+\frac{p\delta}{r}\right),              \eqno(9b)$$
where $\delta$ is defined by (2b). By using Eq. (3) one can
rewrite (9) as
 $$\frac{a_{Z}}{a'}=-\cos\theta'\left(1-\frac{2}{r'}\Delta r
  -\frac{\sin\theta'}{\cos\theta'}\Delta\theta
  +\frac{p\delta}{r'}\right),
                                                 \eqno(10a)$$
  $$\frac{a_{X}}{a'}=-\sin\theta'\left(1-\frac{2}{r'}\Delta r
  +\frac{\cos\theta'}{\sin\theta'}\Delta\theta
  +\frac{p\delta}{r'}\right),                    \eqno(10b)$$
and
  $$ a'= \frac{GM}{r'^2},                      \eqno(10c)$$
where $r'(\theta')$ presents the orbit of the non-rotating outer
shell and $a'$ is its acceleration. We then have the relative
difference of the accelerations for the rotating ball and the
non-rotating shell:
 $$\frac{\eta_{Z}}{\delta} \equiv \frac{2\left(a_{Z}-a'_{Z}\right)}
                                 {\left(a+a'\right)\delta}
   =\cos\theta'\left(\frac{2}{r'}\frac{\Delta r}{\delta}
  +\frac{\sin\theta'}{\cos\theta'}\frac{\Delta\theta}{\delta}
  -\frac{p}{r'}\right),                               \eqno(11a)$$
 $$\frac{\eta_{X}}{\delta} \equiv \frac{2\left(a_{X}-a'_{X}\right)}
                                 {\left(a+a'\right)\delta}
   =\sin\theta'\left(\frac{2}{r'}\frac{\Delta r}{\delta}
  -\frac{\cos\theta'}{\sin\theta'}\frac{\Delta\theta}{\delta}
  -\frac{p}{r'}\right).                              \eqno(11b)$$
The numerical results of $\eta_{Z}$ and $\eta_{Z}$ in unit of
$\delta$ are given in Fig. 2.
\begin{figure}
%\begin{center}
  \leavevmode
 \epsfysize=2.5in
  \epsfbox{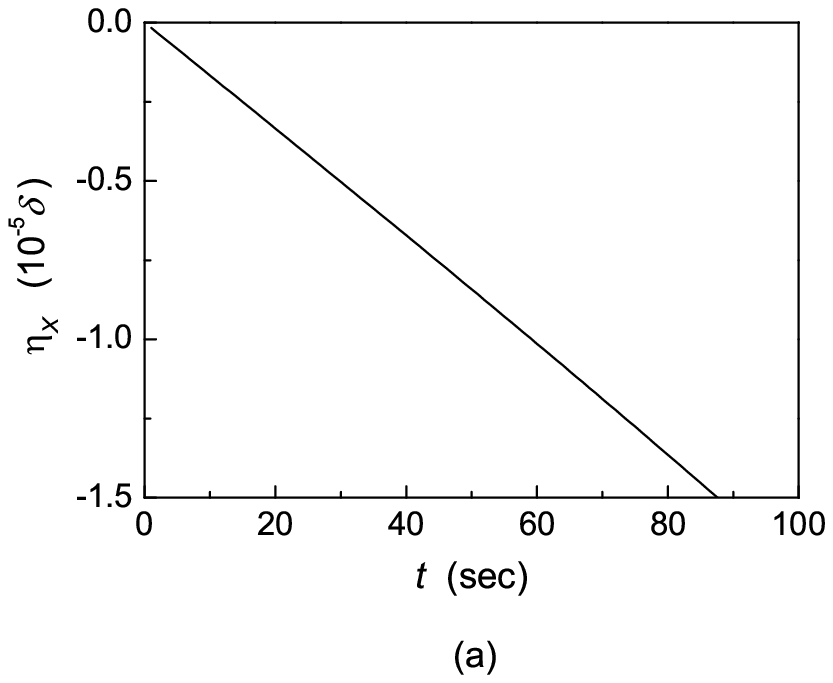}
 \epsfysize=2.5in
  \epsfbox{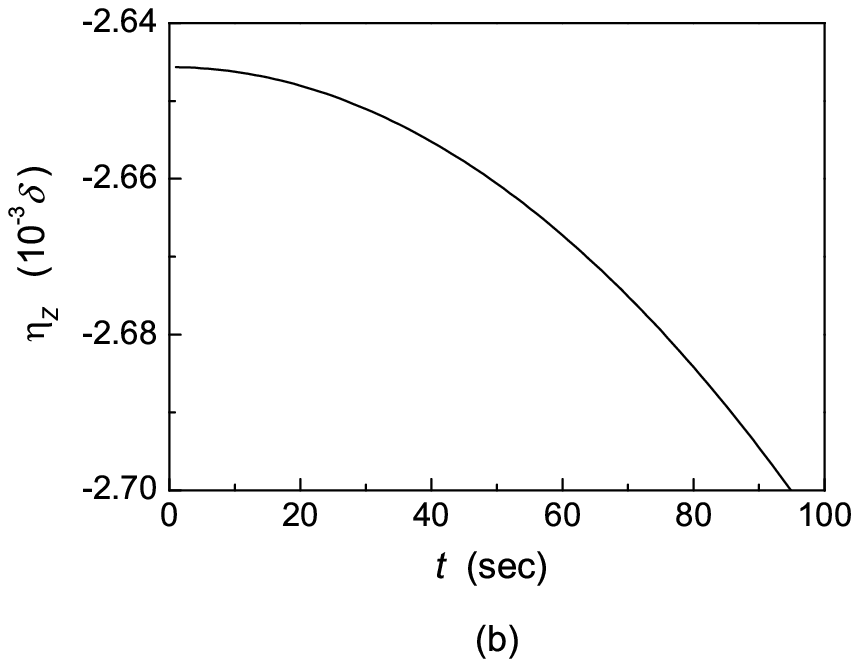}
%\end{center}

 {\footnotesize Fig. 2. (a) and (b) show, respectively, $\eta_X$
 and $\eta_Z$ as functions of time $t$, where  $h=4\times 10^4$ m,
 $R=6.38\times 10^6$ m and $a=\pi /3$ are used. The time $t$ is
 in unit of second. $\eta_X$ and $\eta_Z$ are, respectively,
 in units of $\delta\times 10^{-5}$ and $\delta\times 10^{-3}$.}
%\label{Fig1}
\end{figure}

 In Figs. 1 and 2, the units in the vertical axes depend on the
 quantity $\delta$
 defined by Eq. (2b). For such a rotating ball with the radius of
 5cm and the rotational speed of $4\times 10^4$ rpm, from Eq. (2b)
 where $\kappa\leq 3.4 \times 10^{-31}{\rm gram}^{-1}$(see [15])
 we get
 $$\delta \leq 2\times 10^{-12}.                 \eqno(12)$$

 For the drop capsule of 40 kilometer altitude, the free-fall time
 of the gyroscope is about, for instance, 40 seconds. In this case
 the maximal differences between the rotating inner ball and the
 non-rotating shell can be found from Figs. 1 and 2 and Eq. (12)
 as follows:
  $$\Delta Z \cong 21\delta {\rm m} \leq 4.2\times 10^{-9}
 {\rm cm}, ~~~~ \Delta X \cong 1.7\delta {\rm cm} \leq 0.34
 \times 10^{-11} {\rm cm},                    \eqno(13)$$
 $$\eta_z \cong 2.66\times 10^{-3}\delta\leq 5.3\times 10^{-15},
 ~~~~ \eta_X \cong 0.67\times 10^{-5}\delta\leq 1.3\times
 10^{-17}.                                    \eqno(14)$$

The predictions given by Eqs. (13) and (14) could be detected by a
SQUID magnetometer because such a magnetometer can be used to
measure displacements as small as $10^{-13}$ centimeters.

\vspace{5mm}
 \noindent
  {\large \bf Acknowledgments}
 This project was in part supported by NNSFC under Grant Nos. 19835040,
 10175070 and 10047004 as well as also by NKBRSF G19990754.

\end{document}